\documentclass[12pt]{iopart}
\usepackage{graphicx}
\begin{document}

\title{Open-charm enhancement at FAIR?}

\author{L Tol\'os$^{1,2,3}$, J Schaffner-Bielich$^2$ and H St\"ocker$^{2,3}$}

\address{$^1$ Gesellschaft f\"ur Schwerionenforschung, Planckstrasse 1, 64291 Darmstadt, Germany}
\address{$^2$ Institut f\"ur Theoretische Physik, J.W. Goethe-Universit\"at, Max-von-Laue 1, 60438 Frankfurt (M), Germany} 
\address{$^3$ FIAS, J.W. Goethe-Universit\"at, Max-von-Laue 1, 60438 Frankfurt (M), Germany}

\begin{abstract}
We have calculated the $D$-meson spectral density at finite temperature  within a self-consistent coupled-channel approach that generates dynamically  the $\Lambda_c$ (2593) resonance. We find  a small mass shift for the $D$-meson in this hot and dense medium while the spectral density develops a sizeable width. The reduced attraction felt by the $D$-meson in hot and dense matter together with the large width  observed  have important consequences for the $D$-meson production in the future CBM experiment at FAIR.
\end{abstract}




\vspace{-1cm}

\section{Introduction}

The future CBM experiment (Compressed Baryonic Matter) at the FAIR project (Facility for Antiproton and Ion Research)  at GSI will explore  matter in the region of high-baryon densities and moderature temperatures \cite{fair}. 
Among others, it will address
the in-medium modifications of open-charm mesons. The medium modifications of $D$-mesons have important consequences for $J/\Psi$ suppression \cite{NA501} as well as  open-charm enhancement in nucleus-nucleus collisions \cite{NA50e}. The $J/\Psi$ suppression can be understood in an hadronic environment due to  inelastic comover scattering  and, therefore, the medium modification of
 the $D$-mesons should modify the $J/\Psi$ absorption. On the other hand,
the NA50 Collaboration \cite{NA50e} has observed an enhancement of dimuons in Pb+Pb collisions 
which was tentatively attributed to an open-charm enhancement in A+A collisions by introducting an attractive mass shift for $D$-mesons in the nuclear medium \cite{cassing}. However, the latest results on dimuon production by NA60 \cite{NA60} seem to disregard this possibility. Finally, the $D$-mesic nuclei, predicted by the quark-meson coupling (QMC) model \cite{qmc} are the result of considering an attractive $D$-meson potential.

Calculations based on the QMC model \cite{qmc}, QCD sum-rule (QSR) \cite{arata} and chiral models \cite{amruta} obtain attractive mass shifts  of -50 MeV to -200 MeV at nuclear matter saturation density $\rho_0$, although a second analysis using QSR predicted only  a splitting of $D^+$ and $D^-$ masses of 60 MeV at $\rho_0$ \cite{weise}. In all these investigations, the $D$-meson spectral density in dense matter is not studied. In our previous work \cite{Tolos04}, the $D$-meson spectral density is obtained by including coupled-channel effects as well as the dresssing of the intermediate propagators. Thus, the attractive potential felt by the $D$-meson is strongly reduced or becomes slightly repulsive \cite{Tolos04}, which has been recently supported by \cite{Lutz05}. In this paper, finite temperature effects are included in the determination of the $D$-meson spectral density in order to adapt our calculation to the conditions of density-temperature expected for the CBM experiment \cite{fair}. Our results indicate that the width of the $D$-meson is the only source of open-charm enhancement at FAIR \cite{Tolos06}.

\section{Formalism}
We obtain the  in-medium $D$-meson spectral density at finite temperature taking, as bare interaction, a separable potential which parameters, coupling constant and cutoff, are determined by fixing the position and the width of the $\Lambda_c(2593)$ resonance (see \cite{Tolos04}). Then, the in-medium $DN$ interaction or G-matrix at finite temperature reads
\begin{eqnarray}
&&\hspace{-2cm}\langle M_1 B_1 \mid G(\Omega,T) \mid M_2 B_2 \rangle = \langle M_1 B_1
\mid V \mid M_2 B_2 \rangle  +\nonumber \\
&&\hspace{-2cm}\sum_{M_3 B_3} \langle M_1 B_1 \mid V \mid
M_3 B_3 \rangle \frac {F_{M_3 B_3}(T)}{\Omega-E_{M_3}(T) -E_{B_3}(T)+i\eta} \langle M_3
B_3 \mid
G(\Omega,T)
\mid M_2 B_2 \rangle \ ,
   \label{eq:gmat1}
\end{eqnarray}
where $V$ is the separable potential and $\Omega$ is the  starting energy. In this equation, $M_i$ and $B_i$  represent the possible mesons ($D$,$\pi$,$\eta$) and baryons ($N$,$\Lambda_c$,$\Sigma_c$), respectively. The function $F_{M_3 B_3}(T)$ for the $D N$ states stands for the Pauli operator, i.e  $Q_{D N}(T)=1-n(k_N,T)$, where
$n(k_N,T)$ is the nucleon Fermi distribution at the corresponding
temperature. The function $F_{M_3 B_3}(T)$ is
$1+n(k_{\pi},T)$, with $n(k_{\pi},T)$ being the Bose distribution of pions at a given temperature, for $\pi \Lambda_c$ or $\pi \Sigma_c$ states  while it is unity for the other intermediate states. Furthermore, the properties of the intermediate states are also modified in the medium at finite temperature. For nucleons, we use a temperature-dependent Walecka-type $\sigma-\omega$ model with density-dependent scalar and vector coupling constants \cite{Tolos02}. In the case of pions, the self-energy in nuclear matter at finite temperature is obtained following  the Appendix of \cite{Tolos02}.

The $D$-meson  potential at a given temperature  is then calculated according to
\begin{eqnarray}
\hspace{-2cm} U_{D}(k_{D},E_{D},T)= \int d^3k_N \ n(k_N,T) \ \langle D N \mid
 G_{D N\rightarrow D N} (\Omega = E_N+E_{D},T) \mid D N \rangle \ .
\label{eq:self}
\end{eqnarray}
As for the case of T$=$0, this is a self-consistent problem for the $D$-meson potential, since the in-medium $DN$ interaction depends on the $D$-meson single-particle energy, which in turn depends on the $D$-meson potential.
After achieving self-consistency for the on-shell value
$U_{D}(k_{D},E_D,T)$, we obtain  the
self-energy $\Pi_D(k_D,\omega,T)=2\sqrt{m_D^2+k_D^2} \, U_{D}(k_D,\omega,T)$
and the corresponding spectral density is
\begin{equation}
S_{D}(k_{D},\omega,T) = - \frac {1}{\pi} \mathrm {Im\,} \frac {1}{\omega^2-m_{D}^2-k_{D}^2
-2 \sqrt{m_{D}^2+k_D^2} U_{D}(k_{D},\omega,T)} \ .
\label{eq:spec}
\end{equation}

\begin{figure}[htb]
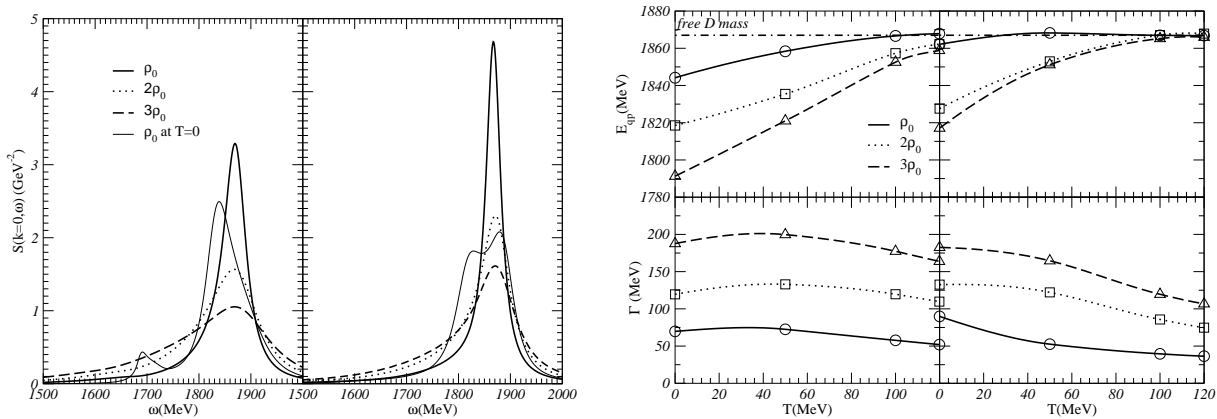

\begin{minipage}[t]{80mm}
     \includegraphics[width=8cm]{tolos1.eps}
\end{minipage}
\begin{minipage}[t]{80mm}
\includegraphics[width=8cm]{tolos2.eps}
\end{minipage}
      \caption{Left: $D$-meson spectral density at $k_D=0$ and T$=$120 MeV as a function of energy for different densities, together with the $D$-meson spectral density at  $k_D=0$ and T$=$0 MeV for $\rho_0$ in the two approaches considered. Right: Quasiparticle energy and width of the $D$-meson spectral density at $k_D=0$ as a function of temperature for different densities and the two approaches considered.}
        \label{fig:dmeson1}
\end{figure}

\section{Results and Conclusion}

In the l.h.s. of Figure \ref{fig:dmeson1} the $D$-meson spectral density at zero momentum and T$=$120 MeV is shown  for different densities and for $\Lambda=1$ GeV and $g^2=13.4$, which reproduce the position and width of the $\Lambda_c(2593)$ resonance (see \cite{Tolos04}). The temperature is chosen in accord with the expected temperatures at FAIR. The spectral density is displayed for the two approaches considered: self-consistent calculation of the $D$-meson self-energy including the dressing of the nucleons in the intermediate states (left panel) and the self-consistent calculation including not only the dressing of nucleons but also the self-energy of pions (right panel). The spectral density at T$=$0 for nuclear matter saturation density,
 $\rho_0=0.17$ fm$^{-3}$, is also shown.
Compared to the T$=$0 case, the quasiparticle peak at finite temperature stays closer to its free position for the range of densities analyzed (from $\rho_0$ up to $3\rho_0$). This is due to the fact the Pauli blocking is reduced with increasing temperature. 
Furthermore, structures present in the spectral distribution at T$=$0 due to the presence of the $\Lambda_c(2593)$ resonance, as reported in \cite{Tolos04}, are washed out. However, the $D$-meson spectral density shows a sizeable width.

Our self-consistent coupled-channel calculation is in stark contrast with previous works based on the QMC model \cite{qmc}, QSR rules \cite{arata} or chiral effective Lagrangians \cite{amruta} which predict a strongly attractive $D$-nucleus potential. We find that the coupled-channel effects at zero temperature result in an important reduction of the in-medium modifications and  are responsible for the considerable width of the $D$-meson, which was not obtained in the previous mean-field works. This effect is independent of the in-medium properties of the intermediate states, as seen in l.h.s of Figure 1. Actually, a recent study of the $D$-meson spectral distribution at T$=$0 suggests a two-mode structure with a repulsive main branch, due to the presence of a new resonance, the $\Sigma_c(2620)$ \cite{Lutz05}. Finite temperatures effects even make the quasiparticle peak get closer to the $D$-meson free mass and $D$-mesons  only show a significant width, as seen in the following.

The r.h.s of Figure \ref{fig:dmeson1} shows the quasiparticle energy together with the width of the $D$-meson spectral density at zero momentum as a function of the temperature for the previous densities and for the approaches considered before. For T$=$0 we observe an attractive potential of -23 MeV for $\rho_0$  and -76 MeV for $3\rho_0$ when $D$-mesons and nucleons are dressed in the intermediate states (upper left panel). For the full self-consistent calculation (upper right panel), the $D$-meson potential at T$=$0 lies between -5 MeV for $\rho_0$ and -48 MeV for $ 3 \rho_0$. For higher temperatures, the quasiparticle peak gets close to the $D$-meson free mass, so there is almost no mass shift expected at finite temperature. On the other hand, the width of the spectral density depends weakly on the temperature. At  T$=$120 MeV the width increases from 52 MeV to 163 MeV for $\rho_0$ to $3\rho_0$ for the first approach (lower left panel) and from 36 MeV at $\rho_0$ to 107 MeV at $3 \rho_0$ for the second approach (lower right panel).

Based on  the previous mean-field calculations which obtain a large $D$-meson mass shift, an enhancement of open-charm in A$+$A collisions  was predicted in order to understand the enhancement of 'intermediate-mass dileptons' in Pb$+$Pb collisions at SPS energies \cite{cassing}. According to our model, the inclusion of a considerable width of the $D$-meson in the medium (40-50 $\rho/\rho_0$ for T$=$120 MeV) is the only source of enhanced in-medium $D$-meson production, as studied for kaons in \cite{Tolosratio}. As a consequence, an off-shell transport theory to account for the $D$-meson production is needed. For that purpose, not only the $D$-meson spectral density but also in-medium $D$-meson cross sections are required. In our model, the cross sections at threshold are expected on the order of 1-20 mb for the range of densities studied in both approaches.

Mesons with charm content at beam energy close to threshold will be investigated by the CBM experiment \cite{fair}. Our results indicate that the mass of the  $D$-meson  is not modified but $D$-mesons show a considerable width in  this hot and dense  medium.

\section*{Acknowledgments}

L.T. acknowledges financial support from AvH Foundation and GSI.


\section*{References}

\begin{thebibliography}{999}

\bibitem{fair} See http://www.gsi.de/fair/experiments/CBM

\bibitem{NA501}
	Gonin A {\it et al.} 1996
	{\it Nucl. Phys.} A {\bf 610} 404c--417c

\bibitem{NA50e}
	Abreu M {\it et al.} 2000 {\it Eur. Phys. J} C {\bf 14}  443--455


\bibitem{cassing}
	Cassing W, Bratkovskaya E L and Sibirtsev A 2001
	{\it Nucl. Phys.} A {\bf 691} 753--778

\bibitem{NA60} Scomparin E, Talk given at QM05 (Budapest, Hungary, 2005)

\bibitem{qmc}
	Tsushima K, Lu D H, Thomas A W, Saito K  and  Landau R H 1999
	{\it Phys. Rev.} C {\bf 59} 2824--2828;
	Sibirtsev A, Tsushima A  and  Thomas A W 1999
	{\it Eur. Phys. J.} A {\bf 6} 351--359

\bibitem{arata}
	Hayashigaki A 2000 {\it Phys. Lett.} B {\bf 487} 96--103

\bibitem{amruta}  Mishra A, Bratkovskaya E L, Schaffner-Bielich J,
Schramm S and  St\"ocker H 2004 {\it Phys. Rev.} C {\bf 69} 015202 1--11

\bibitem{weise}
Weise W 2001 {\it Hirschegg '01: Structure of Hadrons: 29th International Workshop on Gross Properties of Nuclei and Nuclear Excitations} (Hirschegg 2001, Structure of hadrons), p. 249

\bibitem{Tolos04} Tol\'os L,  Schaffner-Bielich  J and Mishra A 2004 {\it Phys. Rev}. C {\bf 70} 025203 1--10

\bibitem{Lutz05}  Lutz M F M and  Korpa  C L 2006 {\it Phys. Lett.} B {\bf 633} 43--48

\bibitem{Tolos06} Tol\'os L, Schaffner-Bielich J and St\"ocker H 2006 {\it Phys. Lett} B {\bf 635} 85--92 

\bibitem{Tolos02} Tol\'os L, Ramos A and Polls A 2002 {\it Phys. Rev.} C {\bf 65} 054907 1--10 

\bibitem{Tolosratio} Tol\'os L, Polls A,  Ramos A and Schaffner-Bielich J 2003 {\it Phys. Rev.} C {\bf 68} 024903 1--11

\end{thebibliography}
\end{document}